\documentclass[submission]{eptcs}
\providecommand{\event}{MARS~2024} 

\usepackage{iftex}
\ifpdf
  \usepackage{underscore}         
  \usepackage[T1]{fontenc}        
\else
  \usepackage{breakurl}           
\fi

\usepackage{xspace}
\usepackage{listings}
\usepackage{cadp-lnt}
\usepackage{cadp-mcl}
\usepackage{cadp-svl}
\usepackage{pss}
\usepackage{tikz}

\usepackage{xcolor}
\definecolor{mymauve}{rgb}{0.1,0.2,0.7}
\definecolor{olivegreen}{cmyk}{.6,.4,0.8,0}

\title{Testing Resource Isolation for System-on-Chip Architectures}

\author{
  Philippe Ledent \qquad\qquad\qquad
  Radu Mateescu \qquad\qquad\qquad
  Wendelin Serwe
  \institute{Univ.~Grenoble Alpes, Inria, CNRS, Grenoble~INP\thanks{Institute of Engineering Univ.~Grenoble Alpes}, LIG, 38000 Grenoble, France}
  \email{
    philippe.ledent@inria.fr \qquad\qquad
    radu.mateescu@inria.fr \qquad\qquad
    wendelin.serwe@inria.fr}
}
\def\titlerunning{Testing Resource Isolation for System-on-Chip Architectures}
\def\authorrunning{Ph.~Ledent, R.~Mateescu \& W.~Serwe}
\begin{document}
\maketitle

\begin{abstract}
  Ensuring resource isolation at the hardware level is a crucial step towards more security inside the Internet of Things.
  Even though there is still no generally accepted technique to generate appropriate tests, it became clear that tests should be generated at the system level.
  In this paper, we illustrate the modeling aspects in test generation for resource isolation, namely modeling the behavior and expressing the intended test scenario.
  We present both aspects using the industrial standard PSS and an academic approach based on conformance testing.
\end{abstract}

\section{Introduction}
\label{sec:introduction}

SoC (System-on-Chip) architectures are being designed and deployed as microcontrollers of embedded systems.
An SoC is usually highly configurable in order to perform several specific tasks in numerous devices, including smartphones or objects in the IoT (Internet of Things).
SoC security is gaining importance as SoCs become ubiquitous, notably because they are being specially manufactured for heavy usage of machine learning and artificial intelligence in the IoT.
Considering the distributed nature of the IoT, software solutions to security are insufficient, because an attacker can easily gain access to some hardware and tamper with it. 
Hardware attacks consist in forcing an SoC to perform operations in order to access functionalities or information that should normally not be available.
A critical security requirement is \emph{resource isolation}, which forbids applications (or programs) running on a same SoC to access data not intended for them.

Ensuring this requirement at hardware level is hence becoming mandatory to strengthen security, but is complex and still leaves two challenging problems.
First, there is yet no commonly accepted solution:~\cite{rodolipin2023} claims to have found a side channel attack that might be applicable to any microcontroller and enable an attacker to access data from secure memory.
Second, there is yet no commonly accepted approach for validating a proposal for a hardware resource isolation solution:
most research focuses on attacking hardware implementations instead of formally validating proposed protocols.

When it comes to IoT devices, microcontroller manufacturers use the ARM Platform Security Architecture\footnote{\url{https://newsroom.arm.com/news/psa-next-steps-toward-a-common-industry-framework-for-secure-iot}} which comes with a security specification and the possibility of certification by ARM (PSA-Certified\footnote{\url{https://www.psacertified.org/}}).
The ARM Security Models~\cite{ARM-PSM} is the open-source ARM architecture for IoT with security concerns.
Here, we focus on the resource isolation aspects of the ARM Security Models that are implemented with the notions of \emph{security} (TrustZone~\cite{TZ}) and \emph{privilege} (TrustZone alone not being enough~\cite{TZresearch}).
ARM provides the possibility to carry security and privilege over the hardware through signals of its AMBA communication protocols~\cite{AMBA} between a source and a target component.
Filtering properly this information can then be left to the target or a dedicated component on the way in charge of monitoring the communication.

Before certifying an SoC by ARM, industrial manufacturers are concerned about representing and testing resource isolation for themselves (the case study~\cite{ARM-PSA-Certified-study} showed that ARM-Certified Level~2 may \emph{leak} confidential information such as AES encryption keys).
Resource isolation should ensure that data contained in an IP (Intellectual Property, as are components usually called in the hardware community) protected with given security and privilege levels can only be accessed by an IP with corresponding or higher levels.
This kind of requirement can be checked using classical tools and techniques for industrial verification, such as hardware simulators using directed tests and/or execution-time assertions.
Although properly written assertions are perfect to monitor exactly the behavior of a design under test during a simulation, it is still necessary to generate appropriate test scenarios to be executed: on its own, assertion-based verification cannot generate such scenarios.

The terrifying complexity of modern SoCs pushes to represent and reason about SoC behavior at higher abstraction levels, to ease the fast generation of many tests.
For this purpose, PSS (the Portable-test Stimulus Standard)~\cite{PSS-2.0} was published by the Accellera Consortium\footnote{\url{https://accellera.org}} that comprises manufacturers such as AMD, ARM, Intel, Nvidia, NXP, and STMicroelectronics, but also major CAD tool vendors, such as Cadence, Siemens EDA, and Synopsis.
PSS aims at providing an easy way to generate (many) tests, without the prior need to explicitly model too much of the SoC's behavior.
PSS defines a (programming) language to abstract the behavior of an SoC as a set of ``\emph{actions}'', which communicate and interact through ``\emph{flow objects}''.
PSS also defines a methodology to generate tests from a VI (``\emph{Verification Intent}'', a test scenario given as a partial ordering of the actions) by filling any gaps of the VI with appropriate actions, meeting the ordering constraints expressed for the SoC.
Industrial manufacturers are inclined to use PSS, because it uses a familiar syntax (close to C++) and is well integrated in their current design flow and tools.

Although PSS has the appearance of a model-based testing approach, the emphasis is clearly more on the test generation, trying to minimize the time spent on the modeling.
Furthermore, because there is no formal semantics of PSS, nor a complete definition of the underlying behavior corresponding to the set of constraints describing an SoC in PSS, the tasks of verification engineers remain difficult.
The major challenge faced by these PSS users is getting a grasp on the behavior used as basis for test generation.
Frequently, an erroneous constraint is only detected when an unexpected test is generated, limiting the confidence in the quality and coverage of the generated tests.

In this paper, we compare the modeling-related aspects of two approaches for test-case generation, namely the PSS approach with an approach based on conformance test generation with test purposes~\cite{Jard-Jeron-05} as supported by the CADP toolbox~\cite{Garavel-Lang-Mateescu-Serwe-13} and its modeling language LNT~\cite{Garavel-Lang-Serwe-17}.
Both approaches involve two separate modeling tasks: coming up with an abstract model of the SoC's behavior and expressing the structure of the desired test scenarios.
However, the focus of both approaches is different: conformance testing starts with a model, whereas PSS favors modeling the test scenarios.
This reflects the needs of verification engineers in the hardware design industry: at the end of the day, they have to produce tests for the SoC, and modeling is is acceptable only if it serves this purpose.
We also study the impact of the difference in focus on the generated test suites.

We illustrate both approaches on the problem of generating tests for resource isolation, using a model of an SoC where the details of the various bus communication protocols are abstracted (each transaction is represented by a single rendezvous), because their differences and details are irrelevant to the test case generation.
For both approaches, we separately discuss the modeling challenges concerning the behavior of the SoC and the structure of the test scenarios.

Formal verification is slowly being integrated in SoC design and verification workflows as shown in survey~\cite{FM-Survey} but not for testing resource isolation.
The closest work to our approach is~\cite{ARM-Intel-Compare} which proposes a high-level model of Intel~64 and ARMv8-A architectures to compare them but it neither formally specifies the behavior nor is the model used as basis for test generation.

The rest of this paper is organized as follows.
Section~\ref{sec:modeling} presents and compares several models of the resource-isolation related SoC behavior in LNT and PSS.
Section~\ref{sec:test} presents the modeling of test scenarios as test purposes in LNT and verification intents in PSS, together with the resulting test suites (sets of generated tests).
Section~\ref{sec:conclusion} concludes.
The complete LNT and PSS code is given in the appendices and provided in the MARS model repository.

\section{Modeling the SoC Behavior for Resource Isolation}
\label{sec:modeling}

We illustrate resource isolation on an SoC with two kinds of IPs (components): sources (e.g., a CPU) and targets (e.g., a memory or dedicated hardware component storing sensible data).
All IPs communicate through a bus-like shared interconnect, which can handle a single transaction at a time.\footnote{There are more complex communication protocols enabling a source to initiate further transactions with other targets, but this requires more than one interconnect.}
Both source and target have a \emph{security level} (secure and non-secure) and a \emph{privilege level} (privileged and non-privileged).
Each target stores a data (data1 or data2).
Each source can execute transactions to read or write the data of a target, or change the security and/or privilege levels of the data stored by the target.
Each transaction consists of an access request emitted by the source, followed by a response (grant or reject) from the target.
Each request by the source to the target includes the security and privilege levels of the source, as is the case for any AMBA~\cite{AMBA} conform hardware protocol.
The target should grant a read or write access if and only if both the security and privilege level of the source are at least those of the target.
Concretely, a read or write request is rejected in the two (non-exclusive) situations where the target is secure (respectively, privileged) and the source is not.
Changing the security and/or privilege level of the target is only granted to a secure and privileged source.
We also allow the source to change its configuration (data to be written and security and privilege levels), thus including the case where a source (CPU) executes applications with different security and privilege levels.

\usepgflibrary{shapes.geometric}
\tikzstyle{state}=[circle,draw=blue!50,fill=blue!20,thick]
\tikzstyle{choice}=[diamond,draw=orange!50,fill=orange!20,thick]
\begin{figure}
  \begin{center}
    \begin{tabular}{c@{\hspace{3em}}c}
    \scalebox{0.79}{\begin{tikzpicture}
  \node (idle) at (0, 0) [state] {};

  \node (Read) at (0, 2) [state] {};
  \draw [thick, ->] (idle) to                  node[fill=white]{\lstinline+!Read+} (Read);
  \draw [thick, ->] (Read) to [out=315, in=45] node[right]{\lstinline+?Reject+} (idle);
  \draw [thick, ->] (Read) to [out=225,in=135] node[left]{\lstinline+?Grant_Read+} (idle);

  \node (Write) at (2.5, -1) [state] {};
  \draw [thick, ->] (idle)  to                  node[fill=white]{\lstinline+!Write+} (Write);
  \draw [thick, ->] (Write) to [out=90, in=0]   node[fill=white,right]{\lstinline+?Grant_Write+} (idle);
  \draw [thick, ->] (Write) to [out=180,in=290] node[fill=white,below]{\lstinline+?Reject+} (idle);

  \node (Protection) at (-2.5, -1) [state] {};
  \draw [thick, ->] (idle) to node[fill=white]{\lstinline+!Protection+} (Protection);
  \draw [thick, ->] (Protection) to [out=90, in=180] node[fill=white,left]{\lstinline+?Grant_Protection+} (idle);
  \draw [thick, ->] (Protection) to [out=0,in=250] node[fill=white,below]{\lstinline+?Reject+}
  (idle);

\end{tikzpicture}}
    &
    \scalebox{0.79}{\begin{tikzpicture}
  \node (idle) at (0, 0) [state] {};

  \node (Read) at (0, 2) [choice] {};
  \draw [thick, ->] (idle) to                  node[fill=white]{\lstinline+?Read+} (Read);
  \draw [thick, ->] (Read) to [out=315, in=45] node[right]{\lstinline+!Reject+} (idle);
  \draw [thick, ->] (Read) to [out=225,in=135] node[left]{\lstinline+!Grant_Read+} (idle);

  \node (Write) at (2.5, -1) [choice] {};
  \draw [thick, ->] (idle)  to                  node[fill=white]{\lstinline+?Write+} (Write);
  \draw [thick, ->] (Write) to [out=90, in=0]   node[fill=white,right]{\lstinline+!Grant_Write+} (idle);
  \draw [thick, ->] (Write) to [out=180,in=290] node[fill=white,below]{\lstinline+!Reject+} (idle);

  \node (Protection) at (-2.5, -1) [choice] {};
  \draw [thick, ->] (idle) to node[fill=white]{\lstinline+?Protection+} (Protection);
  \draw [thick, ->] (Protection) to [out=90, in=180] node[fill=white,left]{\lstinline+!Grant_Protection+} (idle);
  \draw [thick, ->] (Protection) to [out=0,in=250] node[fill=white,below]{\lstinline+!Reject+}
  (idle);

\end{tikzpicture}}
    \end{tabular}
    \caption{Symbolic automata representation of source (left) and target (right) behaviors}
    \label{fig:automata}
  \end{center}
\end{figure}
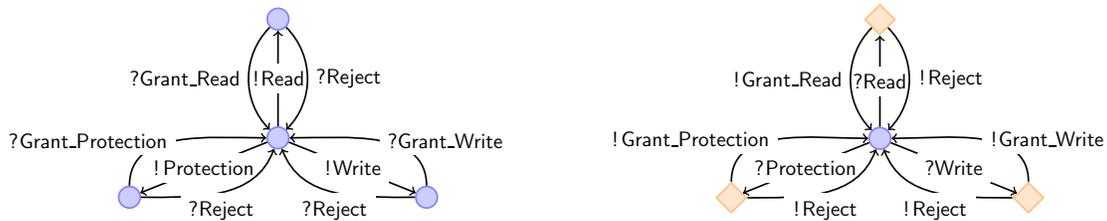

Figure~\ref{fig:automata} shows the behaviors of a source (on the left) and a target (on the right) as two communicating symbolic automata, focusing on the executed actions and hiding all concrete data as well as security and privilege levels.%
\footnote{Taking them into account would yield unreadable figures: for instance, the source automaton would have as many different central (\lstinline+idle+) states as there are different combinations of security and privilege levels (i.e., sixteen).
See also the size of the LTS of the LNT model given in Sect.~\ref{sec:modeling:lnt}.}
Both automata synchronize their transitions with identical labels; we omitted the transition corresponding to a change of the source configuration, because it is the  only unsynchronized transition.

All requests are \emph{initiated} by the source (marked with an exclamation-mark ``!'') and \emph{received} by the target (marked with a question-mark ``?'').
The situation is the opposite for granting or rejecting a request (initiated by the target and received by the source).
Initially, both automata are in their central state, the source is secure and privileged, and the target is non-secure and non-privileged.

The source can attempt a \lstinline+Read+, a \lstinline+Write+, or change the \lstinline+Protection+ level of the target; the target responds by \lstinline+Grant+ing or \lstinline+Reject+ing the request depending of whether or not it was legal.
For each transaction, the source states its security and privilege levels and moves to the state corresponding to its request, where it awaits a response (grant or reject) from the target, before it can issue the next request.

The target starts by awaiting a transaction request from the source.
After reception it enters a state represented by a lozenge symbol indicating that it must analyze the request to determine whether the request will be granted or rejected.
This decision constitutes the resource isolation.
A request is rejected for security reasons if and only if a non-secure source attempts to access a secure target.
Likewise, a request is rejected for privilege reasons if and only if a non-privileged source attempts to access a privileged target.
Altogether, a request is accepted if and only if it is not rejected for either reason.
Only a secure and privileged request can change the security and privilege levels of the target.
Note that the target accepts a new request only in its central state, thus the target accepts a new request only \emph{after} having generated a grant or reject of the pending request (if any).

\subsection{SoC Behavior Modeling in LNT}
\label{sec:modeling:lnt}

\lstset{language=LNT}

Such an SoC can be expressed easily using LNT~\cite{Garavel-Lang-Serwe-17}, a modern language combining a sound foundation in concurrency theory with user-friendly syntax akin to mainstream programming languages.
Two LNT processes define the behavior of a source or target, each encoding the corresponding symbolic automaton as an infinite loop, each iteration of which selects among the various possible actions.
The overall model of the SoC is obtained as a parallel composition of an instance of as many sources and targets as there are in the SoC.
Each communication on the interconnect is modeled as a multiway rendezvous between all instances (reflecting the fact that all IPs can observe everything exchanged on the interconnect).
The complete LNT model (about 200 lines) is given in Appendix~\ref{app:lnt}.

The LNT model defines four enumerated data types: the security (\lstinline+security+) and privilege (\lstinline+privilege+) levels\footnote{Without loss of generality, we restrict the model to two privilege levels (rather than the four considered by ARM).}, the available data values (\lstinline+data+), and the identities of the various IPs (\lstinline+ip+).
For the latter, the model also defines a function \lstinline+source(id)+ returning \lstinline+true+ if and only if \lstinline+id+ identifies a source IP.
The function \lstinline+valid_access(s, t, p, q)+ returns \lstinline+true+ if and only if a source with security level \lstinline+t+ and privilege level \lstinline+q+ should be granted the request to read or write the data stored in a target with security level \lstinline+s+ and privilege level \lstinline+p+. 

\begin{figure}
  \vspace*{-4ex}
\begin{lstlisting}[numbers=left,firstline=98,lastline=150]
process TARGET [Read, Grant_Read, Reject_Read, Write, Grant_Write,
                Reject_Write, Protection, Grant_Protection,
                Reject_Protection: Bus] (id: ip) is
   require not (source (id));
   var d,e: data, s,t,u: security, p,q,r: privilege, o, other: ip in
      d := data1; -- default value
      s := non_secure; p := non_privileged; -- lowest protection level
      loop
         select
            Read (?o, id, ?t, ?q) where source (o);
            if valid_access (s, t, p, q) then
               Grant_Read (o, id, d)
            else
               Reject_Read (o, id)
            end if
         [] Write (?o, id, ?t, ?q, ?e) where source (o);
            if valid_access (s, t, p, q) then
               d := e;
               Grant_Write (o, id)
            else
               Reject_Write (o, id)
            end if
         [] Protection (?o, id, ?t, ?q, ?u, ?r) where source (o);
            if (t == secure) and (q == privileged) then
               s := u; p := r;
               Grant_Protection (o, id, s, p)
            else
               Reject_Protection (o, id)
            end if
         -- communication between other IPs on the shared interconnect
         [] Read (?other, ?o, ?any security, ?any privilege)
               where (o != id) and source (other)
         [] Grant_Read (?other, ?o, ?any data)
               where (o != id) and source (other)
         [] Reject_Read (?other, ?o)
               where (o != id) and source (other)
         [] Write (?other, ?o, ?any security, ?any privilege, ?any data)
               where (o != id) and source (other)
         [] Grant_Write (?other, ?o)
               where (o != id) and source (other)
         [] Reject_Write (?other, ?o)
               where (o != id) and source (other)
         [] Protection (?other, ?o, ?any security, ?any privilege,
                        ?any security, ?any privilege)
               where (o != id) and source (other)
         [] Grant_Protection (?other, ?o, ?any security, ?any privilege)
               where (o != id) and source (other)
         [] Reject_Protection (?other, ?o)
               where (o != id) and source (other)
         end select
      end loop
   end var
end process
\end{lstlisting}
  \vspace*{-6ex}
  \caption{LNT process of a target}
  \label{fig:lnt:target}
\end{figure}

Figure~\ref{fig:lnt:target} shows the LNT process \lstinline+TARGET+ modeling a target.
\lstinline+TARGET+ has a variable parameter \lstinline+id+ identifying the IP; the \lstinline+require+-clause of line~4 enforces that this IP is indeed a target.
Among the ten local variables, three record the currently stored data (\lstinline+d+), security (\lstinline+s+), and privilege (\lstinline+p+).
The other local variables serve to collect values exchanged during the rendezvous, so as to impose constraints (e.g., that the IP emitting a request is a source or whether the request should be granted or rejected based on the security and privilege level of source and target)  or handle data (e.g., change the stored data on line~18 or the security and privilege levels on line~25).
Each request is represented by a rendezvous on the corresponding gate (\lstinline+Read+, \lstinline+Write+, or \lstinline+Protection+), during which the source transmits its current security and privilege levels, which the target stores in its local variables \lstinline+t+ and \lstinline+q+ (this is indicated by the question marks \lstinline[language=LNT]+?+ in lines~10, 16, 23, etc.).
Depending on the validity of the request, the latter is either granted or rejected (by a rendezvous on the corresponding gate).
For a \lstinline+Write+ and \lstinline+Protection+, the grant is preceded by an update of the local variables of the target with the values received from the source during the request (see lines~18 and 25).

The LTS corresponding to a parallel composition of eight sources (which can only initiate the three transactions \lstinline+Read+, \lstinline+Write+ and \lstinline+Protection+) and a single target can be generated in less than a minute, and has 182 states, 558 transitions, and 99 labels (after minimization modulo strong bisimulation).

In a second version of the LNT model, a source not engaged in a transaction can also change its configuration (the data written by the source and the security and privilege level of the source).
This corresponds to considering sources as multitasking-enabled CPUs capable of executing several applications with different configurations, and to take care of the configuration changes induced by switching between applications.
The LTS corresponding to this extended model is too large to be generated---the number of states is expected to be $8^8$ times the size of the previous model.
However, when removing the identification of the source IP from all transition labels and hiding all transitions corresponding to a configuration change, both LTSs are equivalent for branching bisimulation (the LTS minimized modulo branching bisimulation has 52 states, 268 transitions, and 39 labels).

The identity of the source IP seems thus not important.
Indeed, when removing the identification of the source IP from all transition labels and hiding all transitions corresponding to a configuration change, a model with a single multitasking-enabled source also is equivalent for branching bisimulation to the model with eight sources that do not have multitasking enabled.
Hence, with the possibility to change the source configuration, it is sufficient to model a single source.

The situation is more intricate concerning the number of targets.
Actually, two targets are independent and thus equivalent to a single target with two memory cells with separate security and privilege levels.
However, resource isolation is concerned with the access to a single target, so that it is not necessary to study SoCs with more than one target.

It is worth mentioning that the LTS can be analyzed with a full range of verification tools, e.g., those provided by the CADP toolbox.
Besides the equivalence checking tools already used to compare the SoCs with different numbers of sources, it is possible to explore the LTS step by step and to verify temporal logic properties.
This is helpful to gain confidence in the correctness of the modeled behavior.

\subsection{SoC Behavior Modeling in PSS}
\label{sec:modeling:pss}

\lstset{language=PSS}

A major modeling difference between LNT and PSS is that LNT is targeted at modeling the SoC, whereas PSS avoids modeling the overall behavior of the SoC, focusing on simply expressing constraints between the actions of the SoC.
However, the latter is less convenient when it comes to precisely understand the modeled behavior, because it requires to assemble all these constraints together.

The understanding of the behavioral model induced by the constraints can be improved by adopting a modeling discipline, such as encoding the two symbolic automata of Fig.~\ref{fig:automata} (as seen in the previous section, it is sufficient to consider an SoC with a single source and a single target).
For each automaton, each transition can be encoded as a PSS action, which inputs from and outputs to a (same) state flow object storing the data values of the automaton, using constraints to enable actions only for particular states of the automaton and controlling the state resulting from the execution of an action.
Synchronization between the automata is then expressed using stream flow objects, mimicking the multiway rendezvous on the gates in the LNT model.

This intuitive approach yields the PSS model presented in Appendix~\ref{app:pss}, featuring two state flow objects, nine stream flow objects, and a total of 21 actions (ten actions for the transitions of the source, nine actions for the transitions of the target, plus two actions to control the initial state of the two state flow objects---this is required by the PSS semantics).
This significant increase in complexity is accompanied by the need to specify for all actions not only the fields of the state flow object that are modified, but also those that remain unchanged.
All in all, the corresponding PSS model ends up with more than 500 lines.

It is possible to translate this PSS model to LNT (using a translator currently under development), leading to almost two thousand lines of LNT.
This generic translation encodes each action and flow object as a separate LNT process, leading to a total of 32 processes.
The LTS corresponding to each of these processes can be generated and minimized modulo divergence-preserving branching bisimulation, before composing all 32 LTSs into the overall LTS of the PSS model.\footnote{The translation of stream flow objects makes use of the $n$-among-$m$ synchronization currently only supported by the EXP.OPEN~\cite{Lang-05} tool.}
This generation of the corresponding LTS took about a day (on the yeti cluster in the Grenoble site of the Grid'5000 platform), exploiting the 64 cores using a distributed state space generation tool.
However, the corresponding state space (before hiding all transitions related to the interactions between actions and flow objects) is prohibitively large: 1,700,860,640 states, 13,934,786,272 transitions, and 6,706 labels, stored in a file with a size of 88 GB.
Note that more refined compositional generation strategies (e.g., smart generation~\cite{Crouzen-Lang-11}) did not succeed, as some intermediate state spaces for a subset of the processes are larger than the overall state space.

Taking into account that a rendezvous between several actions yields a unique visible transition, we investigated a simpler modeling approach encoding a monolithic automaton, incorporating the constraints of both source and target.
This approach requires only ten actions (three requests, three grants, three rejects, and the configuration change), all inputting from and outputting to a single state flow object.
The corresponding PSS code is given in Appendix~\ref{app:mono}.
The drawback of this approach is the increase in constraints for each action, because it is necessary to specify all fields of the state flow that remain unchanged by the action (each field related to the target is not affected by an action related to the source and vice-versa).
Another inconvenient of this approach is that it would be very impractical to extend this model to an SoC with more IPs, due the complexity of getting a complete and correct set of constraints.

This monolithic PSS model can also be translated into (almost one thousand lines of) LNT, from which the corresponding LTS (2736 states, 4591 transitions, and 4592 labels) can be directly\footnote{Due to the absence of stream flow objects, the generated LNT model does not require a $n$-among-$m$ synchronization and can thus be handled directly by the LNT compiler.} generated in less than a minute.
After hiding all transitions related to interactions with the state flow object, changing all transition labels to use the same gates and sets of offers as the LNT models of the previous section, and determinization (reduction for weak trace equivalence), the LTS is branching equivalent to those of the LNT models presented in the previous section.

\begin{figure}[t]
\begin{lstlisting}[numbers=left,firstline=119,lastline=142]
  action t_grant_read {
    input  system_state in_state;
    output system_state out_state;

    constraint in_state.initial == false;
    // Move from Read to Idle
    constraint in_state.sstate  == read;
    constraint out_state.sstate == idle;
    // Check protection
    constraint (in_state.source_sec == secure) ||
               (in_state.target_sec == non_secure);
    constraint (in_state.source_priv == privileged) ||
               (in_state.target_priv == non_privileged);
    // Maintain source fields
    constraint out_state.source_sec  == in_state.source_sec;
    constraint out_state.source_priv == in_state.source_priv;
    constraint out_state.source_data == in_state.source_data;
    // Maintain target fields
    constraint out_state.target_sec  == in_state.target_sec;
    constraint out_state.target_priv == in_state.target_priv;
    constraint out_state.target_data == in_state.target_data;
    constraint out_state.new_sec     == in_state.new_sec;
    constraint out_state.new_priv    == in_state.new_priv;
  }
\end{lstlisting}
  \caption{Action for granting a read request in the monolithic PSS model}
  \label{fig:pss:grant_read}
\end{figure}

Figure~\ref{fig:pss:grant_read} gives the description of action \lstinline+target_grant_read+.
It inputs from and outputs to a state flow object, which keeps track of the configuration of the SoC.
Execution of the action is subject to the \lstinline+constraint+s specified in its body.
The first constraint (line~5) enforces that the action can be executed only if another action has already output to the state flow object (each PSS state flow object has an implicit field \lstinline+initial+\hspace{-.4em}, which is initialized to \lstinline+true+, changed to \lstinline+false+ upon the first output to the flow object, and never changed again).
The next two constraints express that the source automaton moves from \lstinline+read+ (line~7 constraining the value of field \lstinline+sstate+ of the input flow object \lstinline+in_state+) back to \lstinline+idle+ (line~8 constraining field \lstinline+sstate+ of the output flow object \lstinline+out_state+).
The next two constraints (lines~10--13) express the validity of the transaction (inspecting only fields of the input flow object).
The remaining eight constraints express that all other fields of the output flow object should keep the values of the fields of the input flow object.

This 24-line PSS description of the action (with its constraints) is more verbose than the corresponding three lines of LNT (lines~13--15 in Fig.~\ref{fig:lnt:target}).
This has several reasons.
First, in PSS the states of the target have to be listed explicitly, whereas they are deduced from the control flow in LNT.
Second, LNT has no implicit field \lstinline+initial+\hspace{-.4em}.
Last, but not least, in LNT it is not necessary to specify the variables that maintain their value.

\section{Test Generation from Test Scenarios}
\label{sec:test}

The principal objective of the models of the SoC behavior presented in Section~\ref{sec:modeling} is to enable the generation of tests to validate the SoC.
Characterizing a set of desired tests is a modeling task of its own, based on the idea of expressing a partial ordering of some actions that have to appear in the generated tests, and of relying on tools exploiting the behavioral model to fill in any further actions necessary to obtain a complete test case.
This approach emphasizes the expression of a \emph{test scenario} defining the high-level structure of the tests, leaving the details to automatic tools.
The notion of test scenario is called TP (\emph{test purpose}) in conformance testing theory~\cite{Jard-Jeron-05} and VI (\emph{verification intent}) in PSS.

There are different techniques to construct tests from a test scenario.
The TESTOR tool~\cite{Marsso-Mateescu-Serwe-18} proceeds by a \emph{forward exploration} of a (particular) synchronous product between the TP and the behavioral model, extracting on-the-fly a test or a subgraph called CTG (complete test graph) containing all possible tests for the TP.
The PSS methodology~\cite[Appendix~F]{PSS-2.0} uses a \emph{backward traversal} of the VI, determining for each action its immediately necessary previous actions, based on the constraints in the verification intent and the behavioral model.
In the following, we compare the effect of these different approaches on four test scenarios for resource isolation.

\subsection{Test Scenario 1: Reject for any Reason}
\label{sec:test1}

\begin{figure}
  \begin{center}
    \begin{tabular}{c|c}
    {
\begin{lstlisting}[firstline=129,lastline=140,language=LNT,numbers=left]
process PURPOSE_1 [
           Reject_Read,
           Reject_Write,
           Reject_Protection,
           TESTOR_ACCEPT: none] is
   select
      Reject_Read
   [] Reject_Write
   [] Reject_Protection
   end select;
   loop TESTOR_ACCEPT end loop
end process
\end{lstlisting}}
    &
    {
\begin{lstlisting}[firstline=557,lastline=568,language=PSS,numbers=none]
action intent_1 {
  t_reject_read       Reject_Read;
  t_reject_write      Reject_Write;
  t_reject_protection Reject_Protection;
  activity {
    select{
        Reject_Read;
        Reject_Write;
        Reject_Protection;
    }
  }
}
\end{lstlisting}}
    \end{tabular}
    \caption{Test scenario 1 (``reject for any reason'') as TP in LNT (left) and VI in PSS (right)}
    \label{fig:test1}
  \end{center}
\end{figure}

A natural first test scenario for resource isolation is to search for tests featuring the detection of an illegal transaction, i.e., containing any of the three actions \lstinline+Reject_Read+, \lstinline+Reject_Write+, and \lstinline+Reject_Protection+.
Figure~\ref{fig:test1} shows how to express this scenario as a TP in LNT and a VI in PSS.

\lstset{language=LNT}
In LNT the TP is encapsulated in a process \lstinline+PURPOSE_1+, the gate parameters (lines~2--5) of which are the three actions expected in the scenario plus the special gate \lstinline+TESTOR_ACCEPT+ indicating the goal of the TP.
The behavior of this TP is the sequential composition of a non-deterministic choice (\lstinline+select+ instruction in lines~6--10, choices being separated by ``\lstinline+[]+'') among the three actions, followed by a loop indicating the end of the TP.

\lstset{language=PSS}
In PSS the VI is a compound action, referencing the three actions via action handles (lines~2--4).
The ordering of actions is specified by the \lstinline+activity+ block (lines~5--11), containing a non-deterministic \lstinline+select+ion among the three actions (lines~6--10, choices being separated by ``\lstinline+;+'').

For this TP, TESTOR generates a CTG (183 states, 567 transitions, and 101 labels) that contains all paths to reach any of the three actions, including paths with granted requests before the rejected one.
A CTG can be considered a description of a tester, interacting with the SoC to drive it towards the goal of the TP, by selecting appropriate control actions (or inputs) depending on the outputs observed so far.
In general, a CTG contains states, where the tester has to choose among different control actions to be executed.
The CTG generated for this TP contains 384 choices, all of which can be covered by a suite of 357 test cases that can be generated automatically using the approach proposed in~\cite{Marsso-Mateescu-Serwe-20}.

For this VI, the PSS backward traversal starts by (non-deterministically) choosing one of the three reject actions, and then determines which other actions must immediately precede, by checking which action could have written values to the state flow object so as to satisfy the input constraints of the selected action.
The constraints on the \lstinline+sstate+ field imply the preceding action must be a request.
For \lstinline+Reject_Read+ and \lstinline+Reject_Write+, the constraints on the security and privilege levels imply that in the request, one of these values must be strictly lower than the one of the target.
For the \lstinline+Reject_Protection+, the constraints imply that the preceding \lstinline+Request_Protection+ stems from a source that is not both secure and privileged.
For the monolithic behavioral model, this backward traversal continues until the action \lstinline+init_system_state+ is found.\footnote{For the generic PSS behavioral model, both \lstinline+init_source_state+ and \lstinline+init_target_state+ have to be found.}
In practice, the PSS methodology aims at generating a single test at each invocation.
When implemented using a breadth-first backward traversal (as is the case for some industrial PSS tools), this systematically yields any of the shortest possible tests.

\subsection{Test Scenario 2: Test all Possible Responses (Interleaving Semantics)}
\label{sec:test2}

\begin{figure}
  \begin{center}
    \begin{tabular}{c|c}
      {
\begin{lstlisting}[firstline=145,lastline=162,language=LNT,numbers=left]
process PURPOSE_2 [
           Reject_Read,
           Reject_Write,
           Reject_Protection,
           Grant_Read,
           Grant_Write,
           Grant_Protection,
           TESTOR_ACCEPT: none] is
   par      
      Grant_Read
   || Grant_Write
   || Grant_Protection
   || Reject_Read
   || Reject_Write
   || Reject_Protection
   end par;
   loop TESTOR_ACCEPT end loop
end process
\end{lstlisting}}
    &
    {
\begin{lstlisting}[firstline=571,lastline=588,language=PSS,numbers=none]
action intent_2 {
  t_grant_read        Grant_Read;
  t_grant_write       Grant_Write;
  t_grant_protection  Grant_Protection;
  t_reject_read       Reject_Read;
  t_reject_write      Reject_Write;
  t_reject_protection Reject_Protection;
  activity {
    schedule{
        Grant_Read;
        Grant_Write;
        Grant_Protection;
        Reject_Read;
        Reject_Write;
        Reject_Protection;
    }
  }
}
\end{lstlisting}}
    \end{tabular}
    \caption{Test scenario 2 (``all possible responses'' interleaved) as TP in LNT (left) and VI in PSS (right)}
    \label{fig:test2}
  \end{center}
\end{figure}

This test scenario aims at observing all responses to the three transactions, in any order using the \emph{interleaving} of the responses as shown in Figure~\ref{fig:test2}.
\lstset{language=LNT}
In LNT, the parallel composition operator \lstinline+par+ expresses the interleaving of the different branches separated by ``\lstinline+||+''.
\lstset{language=PSS}
In PSS, the \lstinline+schedule+ operator expresses the interleaving of the branches separated by ``\lstinline+;+''.%
\footnote{The PSS operator \lstinline+parallel+ expresses a parallel execution of different behaviors using several threads.}

For this TP, TESTOR computes a CTG with 2649 states and 12,057 transitions; its 8832 choices can be covered with 8328 tests.
The size of the CTG is due to the fact that once one of the responses has been observed, it is still possible to observe it before all responses have been observed.
Hence, the CTG corresponds to an ``unfolding'' of the model six times, repeating the complete behavior of the SoC until all responses have been observed.

Searching for short(est) tests, the PSS methodology reduces the number of changes in the security and privilege levels of the source and the target.
Therefore, in most tests the security and privilege levels for \lstinline+Grant_Read+ and \lstinline+Grant_Write+ (respectively \lstinline+Reject_Read+ and \lstinline+Reject_Write+) are the same, and \lstinline+Grant_Protection+ and \lstinline+Reject_Protection+ are inserted where suitable.
Notice that the syntactic order of the responses in the VI (and TP) actually corresponds to the shortest sequence.
Indeed, because the model starts with a secure and privileged source and a non-secure and non-privileged target, all grants are possible.
Increasing the security and/or privilege of the target and appropriately lowering the security and privilege of the source are then sufficient to observe the three rejections.

\subsection{Test Scenario 3: Test all Possible Responses (Sequential Semantics)}
\label{sec:test3}

\begin{figure}
  \begin{center}
    \begin{tabular}{c|c}
      {
\begin{lstlisting}[firstline=167,lastline=182,language=LNT,numbers=left]
process PURPOSE_3 [
           Reject_Read,
           Reject_Write,
           Reject_Protection,
           Grant_Read,
           Grant_Write,
           Grant_Protection,
           TESTOR_ACCEPT: none] is
   Grant_Read;
   Grant_Write;
   Grant_Protection;
   Reject_Read;
   Reject_Write;
   Reject_Protection;
   loop TESTOR_ACCEPT end loop
end process
\end{lstlisting}}
      &
      {
\begin{lstlisting}[firstline=591,lastline=606,language=PSS,numbers=none]
action intent_3 {
  t_grant_read        Grant_Read;
  t_grant_write       Grant_Write;
  t_grant_protection  Grant_Protection;
  t_reject_read       Reject_Read;
  t_reject_write      Reject_Write;
  t_reject_protection Reject_Protection;
  activity {
    Grant_Read;
    Grant_Write;
    Grant_Protection;
    Reject_Read;
    Reject_Write;
    Reject_Protection;
  }
}
\end{lstlisting}}
    \end{tabular}
    \caption{Test scenario 3 (``all possible responses'' in sequence) as TP in LNT (left) and VI in PSS (right)}
    \label{fig:test3}
  \end{center}
\end{figure}

Most test generation strategies do not support the interleaving of actions, but require more \emph{directed} specifications enforcing a particular sequence of actions.
Test scenario~3 requests once again all possible responses but in a particular order, expressed in LNT and PSS using ``\lstinline+;+'', as illustrated on Figure~\ref{fig:test3}.

Requesting such a \emph{directed} scenario has consequences on the generated test suite for both LNT and PSS.
The CTG generated by TESTOR will contain for two sequential actions of the TP every possible path of the model allowed in between.
The CTG has 967 states and 3271 transitions; its 2208 choices can be covered with 2072 tests.
This CTG is smaller than the one for test scenario~2, because only a single ordering of responses is requested.

The tests generated by PSS are once again the shortest ones and included in those generated for test scenario~2.
This shows that more directed test scenarios limit the set of generated tests.

\subsection {Test Scenario 4: Access Data with Different Protection}
\label{sec:test4}

Using the notions of security and privilege, ARM-PSA diversifies the different levels of protection possible for an IP in an SoC.
However, there is the strong assumption of a \emph{trusted administrator} as all requests of a secure and privileged source are necessarily granted.
Test scenario~4 expresses that whatever the security and privilege of the target, a source with the same security and privilege can write to the target, and any source with higher security and/or privilege (e.g., the administrator) will be able to read the written data.
This scenario requires to express that there should be no change in the security or privilege between the write and read requests.

\begin{figure}
  \begin{center}
    {
\begin{lstlisting}[firstline=186,lastline=201,language=LNT,numbers=left]
process PURPOSE_4 [Read, Grant_Read, Write, Grant_Protection: Bus,
                   TESTOR_ACCEPT, TESTOR_REFUSE: none] is
   var s,t: security, p,q: privilege, d: data in
      Grant_Protection (?any ip, ip0, ?s, ?p)
      Write (?any ip, ip0, s, p, ?d); -- same s and p as in the previous line
      select
	 -- refuse any further rendezvous on gate Grant\_Protection
         Grant_Protection (?any ip, ip0, ?s, ?p); loop TESTOR_REFUSE end loop
      [] -- accept all other rendezvous
	 null
      end select;
      Read (?any ip, ip0, ?t, ?q) where (s != t) or (p != q);
      Grant_Read (?any ip, ip0, d); -- access data with different security and privilege levels
      loop TESTOR_ACCEPT end loop
   end var
end process
\end{lstlisting}}
    \caption{Test scenario 4 (``access data with different security/privilege'') as TP in LNT}
    \label{fig:test4}
  \end{center}
\end{figure}

Test Scenario~4 focuses on how to express the refusal of some behavior.
This is illustrated in Figure~\ref{fig:test4} by describing a corresponding TP in LNT, using the special gate \lstinline+TESTOR_REFUSE+ (line~8) to indicate that the preceding rendezvous on gate \lstinline+Grant_Protection+ should be excluded from the generated CTG.
The \lstinline+null+ branch (line~10) of the \lstinline+select+ construct (lines~6--11) allows any other action.
The \lstinline+where+ clause on line~12 guarantees (in combination with the condition on line~11 of Figure~\ref{fig:lnt:target}) that the final read is requested with higher security and/or privilege than the write on line~5.

To the best of our knowledge, PSS has no such means to explicitly request absence of actions from the generated tests.
Instead, the scenario has to be made more directed by explicitly including more actions in the VI so as to add constraints on these actions.
In particular, the VI allows to \lstinline+bind+ an input flow object of an action $a_2$ to the output flow object of another action $a_1$, constraining action-inference and forcing $a_1$ to immediately precede $a_2$.
The resulting, lengthy VI is given in Appendix~\ref{app:mono}.

\section{Conclusion}
\label{sec:conclusion}

In this paper, we illustrated the modeling tasks for testing hardware resource isolation using both the approach promoted by the industrial standard PSS and an academic approach based on LNT and conformance testing.
Both approaches require a model of behavior and an abstract test scenario, which is refined into concrete tests based on the behavioral model.

Despite these similarities, both approaches differ in the way of generating tests, using a forward (LNT) or backward (PSS) search.
This difference not only yields different tests, but also impacts the modeling, due to the trade-off between putting constraints in the behavior model or the test scenario.
On the one hand, LNT facilitates a complete, verifiable model of the behavior, from which extensive test suites can be generated with few, short test scenarios.
On the other hand, PSS favors focusing on the test scenario (or verification intent), and requires longer test scenarios to obtain longer tests.
While this avoids the risk of state space explosion, it comes at the price of losing the coverage guarantees available for conformance testing, in particular in the presence of cyclic behavior.
Furthermore, the behavior is often under-constrained in PSS, especially when adding a new action to the behavior.

The models presented in this paper were used in an industrial context.
An extended version of test scenario~3 requested in LNT a specific order of attempting each transaction for all combinations of source and target security and privilege levels.
Concretely, for each attempted transaction, the source requests to write, to read, and then to change the target's security and privilege.
From the generated CTG, we derived a single long test (including all transaction attempts).
This test was included in the nightly non-regression tests for a (confidential) SoC under development, sequentially executing the test for each of the over hundred target IPs of the SoC.
This revealed a few cases of bad wiring, unaligned documentation, and misunderstandings between architect, design, and verification engineers.

Because the behavioral model of PSS is hard to grasp, modeling errors are frequently detected only by the generation of unexpected tests.
We are currently working on the automated translation of PSS constructs into LNT to support the early analysis of the behavioral model, e.g., by model checking.
This also includes guidelines for devising PSS models with an efficient translation to LNT.

\bibliographystyle{eptcs}
\bibliography{mars}

\begin{thebibliography}{10}
\providecommand{\bibitemdeclare}[2]{}
\providecommand{\surnamestart}{}
\providecommand{\surnameend}{}
\providecommand{\urlprefix}{Available at }
\providecommand{\url}[1]{\texttt{#1}}
\providecommand{\href}[2]{\texttt{#2}}
\providecommand{\urlalt}[2]{\href{#1}{#2}}
\providecommand{\doi}[1]{doi:\urlalt{https://doi.org/#1}{#1}}
\providecommand{\eprint}[1]{arXiv:\urlalt{https://arxiv.org/abs/#1}{#1}}
\providecommand{\bibinfo}[2]{#2}

\bibitemdeclare{misc}{AMBA}
\bibitem{AMBA}
\bibinfo{author}{\surnamestart ARM\surnameend}: \emph{\bibinfo{title}{AMBA
  Specification (Rev 2.0)}}.
\newblock \urlprefix\url{https://developer.arm.com/documentation/ihi0011/a}.

\bibitemdeclare{misc}{ARM-PSM}
\bibitem{ARM-PSM}
\bibinfo{author}{\surnamestart ARM\surnameend}: \emph{\bibinfo{title}{Platform
  Security Model 1.1}}.
\newblock
  \urlprefix\url{https://developer.arm.com/documentation/den0128/latest}.

\bibitemdeclare{misc}{TZ}
\bibitem{TZ}
\bibinfo{author}{\surnamestart ARM\surnameend}: \emph{\bibinfo{title}{Security
  in an ARMv8 System}}.
\newblock
  \urlprefix\url{https://developer.arm.com/documentation/100935/0100/Security-in-ARMv8-A-systems-}.

\bibitemdeclare{inproceedings}{ARM-Intel-Compare}
\bibitem{ARM-Intel-Compare}
\bibinfo{author}{Guillaume \surnamestart Averlant\surnameend},
  \bibinfo{author}{Beno\^it \surnamestart Morgan\surnameend},
  \bibinfo{author}{\'Eric \surnamestart Alata\surnameend},
  \bibinfo{author}{Vincent \surnamestart Nicomette\surnameend} \&
  \bibinfo{author}{Mohamed \surnamestart Ka\^aniche\surnameend}
  (\bibinfo{year}{2017}): \emph{\bibinfo{title}{An Abstraction Model and a
  Comparative Analysis of Intel and ARM Hardware Isolation Mechanisms}}.
\newblock In: {\slshape \bibinfo{booktitle}{2017 IEEE 22nd Pacific Rim
  International Symposium on Dependable Computing (PRDC)}}, pp.
  \bibinfo{pages}{245--254}, \doi{10.1109/PRDC.2017.48}.

\bibitemdeclare{article}{ARM-PSA-Certified-study}
\bibitem{ARM-PSA-Certified-study}
\bibinfo{author}{Fei \surnamestart Chen\surnameend}, \bibinfo{author}{Duming
  \surnamestart Luo\surnameend}, \bibinfo{author}{Jianqiang \surnamestart
  Li\surnameend}, \bibinfo{author}{Victor C.~M. \surnamestart
  Leung\surnameend}, \bibinfo{author}{Shiqi \surnamestart Li\surnameend} \&
  \bibinfo{author}{Junfeng \surnamestart Fan\surnameend}
  (\bibinfo{year}{2023}): \emph{\bibinfo{title}{Arm PSA-Certified IoT Chip
  Security: A Case Study}}.
\newblock {\slshape \bibinfo{journal}{Tsinghua Science and Technology}}
  \bibinfo{volume}{28}(\bibinfo{number}{2}), pp. \bibinfo{pages}{244--257},
  \doi{10.26599/TST.2021.9010094}.

\bibitemdeclare{inproceedings}{Crouzen-Lang-11}
\bibitem{Crouzen-Lang-11}
\bibinfo{author}{Pepijn \surnamestart Crouzen\surnameend} \&
  \bibinfo{author}{Fr\'ed\'eric \surnamestart Lang\surnameend}
  (\bibinfo{year}{2011}): \emph{\bibinfo{title}{{Smart Reduction}}}.
\newblock In \bibinfo{editor}{Dimitra \surnamestart Giannakopoulou\surnameend}
  \& \bibinfo{editor}{Fernando \surnamestart Orejas\surnameend}, editors:
  {\slshape \bibinfo{booktitle}{Proceedings of Fundamental Approaches to
  Software Engineering (FASE'11), Saarbr\"ucken, Germany}}, {\slshape
  \bibinfo{series}{Lecture Notes in Computer Science}} \bibinfo{volume}{6603},
  \bibinfo{publisher}{Springer}, pp. \bibinfo{pages}{111--126},
  \doi{10.1007/978-3-642-19811-3_9}.

\bibitemdeclare{article}{Garavel-Lang-Mateescu-Serwe-13}
\bibitem{Garavel-Lang-Mateescu-Serwe-13}
\bibinfo{author}{Hubert \surnamestart Garavel\surnameend},
  \bibinfo{author}{Fr\'ed\'eric \surnamestart Lang\surnameend},
  \bibinfo{author}{Radu \surnamestart Mateescu\surnameend} \&
  \bibinfo{author}{Wendelin \surnamestart Serwe\surnameend}
  (\bibinfo{year}{2013}): \emph{\bibinfo{title}{{CADP 2011: A Toolbox for the
  Construction and Analysis of Distributed Processes}}}.
\newblock {\slshape \bibinfo{journal}{Springer International Journal on
  Software Tools for Technology Transfer (STTT)}}
  \bibinfo{volume}{15}(\bibinfo{number}{2}), pp. \bibinfo{pages}{89--107},
  \doi{10.1007/s10009-012-0244-z}.

\bibitemdeclare{inproceedings}{Garavel-Lang-Serwe-17}
\bibitem{Garavel-Lang-Serwe-17}
\bibinfo{author}{Hubert \surnamestart Garavel\surnameend},
  \bibinfo{author}{Fr\'ed\'eric \surnamestart Lang\surnameend} \&
  \bibinfo{author}{Wendelin \surnamestart Serwe\surnameend}
  (\bibinfo{year}{2017}): \emph{\bibinfo{title}{{From LOTOS to LNT}}}.
\newblock In \bibinfo{editor}{Joost-Pieter \surnamestart Katoen\surnameend},
  \bibinfo{editor}{Rom \surnamestart Langerak\surnameend} \&
  \bibinfo{editor}{Arend \surnamestart Rensink\surnameend}, editors: {\slshape
  \bibinfo{booktitle}{ModelEd, TestEd, TrustEd -- Essays Dedicated to Ed
  Brinksma on the Occasion of His 60th Birthday}}, {\slshape
  \bibinfo{series}{Lecture Notes in Computer Science}} \bibinfo{volume}{10500},
  \bibinfo{publisher}{Springer}, pp. \bibinfo{pages}{3--26},
  \doi{10.1007/978-3-319-68270-9_1}.

\bibitemdeclare{article}{FM-Survey}
\bibitem{FM-Survey}
\bibinfo{author}{Tom\'as \surnamestart Grimm\surnameend},
  \bibinfo{author}{Djones \surnamestart Lettnin\surnameend} \&
  \bibinfo{author}{Michael \surnamestart H\"ubner\surnameend}
  (\bibinfo{year}{2018}): \emph{\bibinfo{title}{A Survey on Formal Verification
  Techniques for Safety-Critical Systems-on-Chip}}.
\newblock {\slshape \bibinfo{journal}{Electronics}}
  \bibinfo{volume}{7}(\bibinfo{number}{6}), \doi{10.3390/electronics7060081}.

\bibitemdeclare{article}{Jard-Jeron-05}
\bibitem{Jard-Jeron-05}
\bibinfo{author}{Claude \surnamestart Jard\surnameend} \&
  \bibinfo{author}{Thierry \surnamestart J\'eron\surnameend}
  (\bibinfo{year}{2005}): \emph{\bibinfo{title}{TGV: Theory, Principles and
  Algorithms -- A Tool for the Automatic Synthesis of Conformance Test Cases
  for Non-Deterministic Reactive Systems}}.
\newblock {\slshape \bibinfo{journal}{Springer International Journal on
  Software Tools for Technology Transfer (STTT)}}
  \bibinfo{volume}{7}(\bibinfo{number}{4}), pp. \bibinfo{pages}{297--315},
  \doi{10.1007/s10009-004-0153-x}.

\bibitemdeclare{inproceedings}{Lang-05}
\bibitem{Lang-05}
\bibinfo{author}{Fr\'ed\'eric \surnamestart Lang\surnameend}
  (\bibinfo{year}{2005}): \emph{\bibinfo{title}{{EXP.OPEN 2.0: A Flexible Tool
  Integrating Partial Order, Compositional, and On-the-fly Verification
  Methods}}}.
\newblock In \bibinfo{editor}{Judi \surnamestart Romijn\surnameend},
  \bibinfo{editor}{Graeme \surnamestart Smith\surnameend} \&
  \bibinfo{editor}{Jaco \surnamestart {van de Pol}\surnameend}, editors:
  {\slshape \bibinfo{booktitle}{Proceedings of the 5th International Conference
  on Integrated Formal Methods (IFM'05), Eindhoven, The Netherlands}},
  {\slshape \bibinfo{series}{Lecture Notes in Computer Science}}
  \bibinfo{volume}{3771}, \bibinfo{publisher}{Springer}, pp.
  \bibinfo{pages}{70--88}, \doi{10.1007/11589976_6}.
\newblock \bibinfo{note}{Full version available as INRIA Research
  Report~RR-5673}.

\bibitemdeclare{article}{TZresearch}
\bibitem{TZresearch}
\bibinfo{author}{Wenhao \surnamestart Li\surnameend}, \bibinfo{author}{Yubin
  \surnamestart Xia\surnameend} \& \bibinfo{author}{Haibo \surnamestart
  Chen\surnameend} (\bibinfo{year}{2019}): \emph{\bibinfo{title}{Research on
  ARM TrustZone}}.
\newblock {\slshape \bibinfo{journal}{GetMobile: Mobile Comp. and Comm.}}
  \bibinfo{volume}{22}(\bibinfo{number}{3}), pp. \bibinfo{pages}{17--22},
  \doi{10.1145/3308755.3308761}.

\bibitemdeclare{inproceedings}{Marsso-Mateescu-Serwe-18}
\bibitem{Marsso-Mateescu-Serwe-18}
\bibinfo{author}{Lina \surnamestart Marsso\surnameend}, \bibinfo{author}{Radu
  \surnamestart Mateescu\surnameend} \& \bibinfo{author}{Wendelin \surnamestart
  Serwe\surnameend} (\bibinfo{year}{2018}): \emph{\bibinfo{title}{{TESTOR: A
  Modular Tool for On-the-Fly Conformance Test Case Generation}}}.
\newblock In \bibinfo{editor}{Dirk \surnamestart Beyer\surnameend} \&
  \bibinfo{editor}{Marieke \surnamestart Huisman\surnameend}, editors:
  {\slshape \bibinfo{booktitle}{Proceedings of the 24th International
  Conference on Tools and Algorithms for the Construction and Analysis of
  Systems (TACAS'18), Thessaloniki, Greece}}, {\slshape
  \bibinfo{series}{Lecture Notes in Computer Science}} \bibinfo{volume}{10806},
  \bibinfo{publisher}{Springer}, pp. \bibinfo{pages}{211--228},
  \doi{10.1007/978-3-319-89963-3_13}.

\bibitemdeclare{inproceedings}{Marsso-Mateescu-Serwe-20}
\bibitem{Marsso-Mateescu-Serwe-20}
\bibinfo{author}{Lina \surnamestart Marsso\surnameend}, \bibinfo{author}{Radu
  \surnamestart Mateescu\surnameend} \& \bibinfo{author}{Wendelin \surnamestart
  Serwe\surnameend} (\bibinfo{year}{2020}): \emph{\bibinfo{title}{{Automated
  Transition Coverage in Behavioural Conformance Testing}}}.
\newblock In: {\slshape \bibinfo{booktitle}{32nd IFIP Int. Conference on
  Testing Software and Systems (ICTSS'20), Naples, Italy}},
  \bibinfo{publisher}{Springer}, pp. \bibinfo{pages}{219--235},
  \doi{10.1007/978-3-030-64881-7\_14}.

\bibitemdeclare{techreport}{PSS-2.0}
\bibitem{PSS-2.0}
\bibinfo{author}{\surnamestart {Portable Stimulus Working Group}\surnameend}
  (\bibinfo{year}{2001}): \emph{\bibinfo{title}{Portable Test and Stimulus
  Standard 2.0}}.
\newblock \bibinfo{type}{Accellera standards}, \bibinfo{institution}{Accellera
  Systems Initiative}, \bibinfo{address}{Elk Grove, CA, USA}.
\newblock
  \urlprefix\url{https://accellera.org/images/downloads/standards/Portable_Test_Stimulus_Standard_v20.pdf}.

\bibitemdeclare{inproceedings}{rodolipin2023}
\bibitem{rodolipin2023}
\bibinfo{author}{C.~\surnamestart Rodrigues\surnameend},
  \bibinfo{author}{D.~\surnamestart Oliveira\surnameend} \&
  \bibinfo{author}{S.~\surnamestart Pinto\surnameend} (\bibinfo{year}{2024}):
  \emph{\bibinfo{title}{BUSted!!! Microarchitectural Side-Channel Attacks on
  the MCU Bus Interconnect}}.
\newblock In: {\slshape \bibinfo{booktitle}{2024 IEEE Symposium on Security and
  Privacy (SP)}}, \bibinfo{publisher}{IEEE Computer Society},
  \bibinfo{address}{Los Alamitos, CA, USA}, \doi{10.1109/SP54263.2024.00062}.

\end{thebibliography}
\appendix
\clearpage

\section{LNT Model}
\label{app:lnt}

\begin{lstlisting}[language=LNT]
module model_8_1 with ==, != is

------------------------------------------------------------------------------

type security is secure, non_secure end type
type privilege is privileged, non_privileged end type

type data is data1, data2 end type

type ip is ip0, ip1, ip2, ip3, ip4, ip5, ip6, ip7, ip8 end type

------------------------------------------------------------------------------

channel Bus is
   (source, target: ip),
   (source, target: ip, d: data),
   (source, target: ip, s: security, p: privilege),
   (source, target: ip, s: security, p: privilege, d: data),
   (source, target: ip, s: security, p: privilege, t: security, q: privilege)
end channel

------------------------------------------------------------------------------

function valid_access (s,t: security, p,q: privilege): Bool is
   -- returns true iff a source with protection level (t,q) is
   -- allowed to access a target with protection level (s,p)
   return not (((s == secure)     and (t == non_secure))      or
               ((p == privileged) and (q == non_privileged)))
end function

------------------------------------------------------------------------------

function source (id: ip) : bool is
   -- returns true iff id is a source IP
   case id in
      ip0 -> return false
   |  any -> return true
   end case
end function

------------------------------------------------------------------------------

process SOURCE [Read, Grant_Read, Reject_Read, Write, Grant_Write,
                Reject_Write, Protection, Grant_Protection,
                Reject_Protection: Bus, Change_Source_Config: any]
               (id: ip, in var s: security, in var p: privilege,
                in var d: data, multitasking: Bool) is
   require source (id);
   var o, other: ip in
      loop
         select
            Read (id, ?o, s, p) where not (source (o));
            select
               Grant_Read (id, o, ?any data)
            [] Reject_Read (id, o)
            end select
         [] Write (id, ?o, s, p, d) where not (source (o));
            select
               Grant_Write (id, o)
            [] Reject_Write (id, o)
            end select
         [] Protection (id, ?o, s, p, ?any security, ?any privilege)
               where not (source (o));
            select
               Grant_Protection (id, o, ?any security, ?any privilege)
            [] Reject_Protection (id, o)
            end select
         [] only if multitasking then
               Change_Source_Config (id, id, ?s, ?p, ?d)
            end if
         -- communication between other IPs on the shared interconnectkbu
         [] Read (?o, ?other, ?any security, ?any privilege)
               where (o != id) and not (source (other))
         [] Grant_Read (?o, ?other, ?any data)
               where (o != id) and not (source (other))
         [] Reject_Read (?o, ?other)
               where (o != id) and not (source (other))
         [] Write (?o, ?other, ?any security, ?any privilege, ?any data)
               where (o != id) and not (source (other))
         [] Grant_Write (?o, ?other)
               where (o != id) and not (source (other))
         [] Reject_Write (?o, ?other)
               where (o != id) and not (source (other))
         [] Protection (?o, ?other, ?any security, ?any privilege,
               ?any security, ?any privilege)
               where (o != id) and not (source (other))
         [] Grant_Protection (?o, ?other, ?any security, ?any privilege)
               where (o != id) and not (source (other))
         [] Reject_Protection (?o, ?other)
               where (o != id) and not (source (other))
         end select
      end loop
   end var
end process

------------------------------------------------------------------------------

process TARGET [Read, Grant_Read, Reject_Read, Write, Grant_Write,
                Reject_Write, Protection, Grant_Protection,
                Reject_Protection: Bus] (id: ip) is
   require not (source (id));
   var d,e: data, s,t,u: security, p,q,r: privilege, o, other: ip in
      d := data1; -- default value
      s := non_secure; p := non_privileged; -- lowest protection level
      loop
         select
            Read (?o, id, ?t, ?q) where source (o);
            if valid_access (s, t, p, q) then
               Grant_Read (o, id, d)
            else
               Reject_Read (o, id)
            end if
         [] Write (?o, id, ?t, ?q, ?e) where source (o);
            if valid_access (s, t, p, q) then
               d := e;
               Grant_Write (o, id)
            else
               Reject_Write (o, id)
            end if
         [] Protection (?o, id, ?t, ?q, ?u, ?r) where source (o);
            if (t == secure) and (q == privileged) then
               s := u; p := r;
               Grant_Protection (o, id, s, p)
            else
               Reject_Protection (o, id)
            end if
         -- communication between other IPs on the shared interconnect
         [] Read (?other, ?o, ?any security, ?any privilege)
               where (o != id) and source (other)
         [] Grant_Read (?other, ?o, ?any data)
               where (o != id) and source (other)
         [] Reject_Read (?other, ?o)
               where (o != id) and source (other)
         [] Write (?other, ?o, ?any security, ?any privilege, ?any data)
               where (o != id) and source (other)
         [] Grant_Write (?other, ?o)
               where (o != id) and source (other)
         [] Reject_Write (?other, ?o)
               where (o != id) and source (other)
         [] Protection (?other, ?o, ?any security, ?any privilege,
                        ?any security, ?any privilege)
               where (o != id) and source (other)
         [] Grant_Protection (?other, ?o, ?any security, ?any privilege)
               where (o != id) and source (other)
         [] Reject_Protection (?other, ?o)
               where (o != id) and source (other)
         end select
      end loop
   end var
end process

------------------------------------------------------------------------------

process SOC [Read, Grant_Read, Reject_Read, Write, Grant_Write, Reject_Write,
             Protection, Grant_Protection, Reject_Protection: Bus,
             Change_Source_Config: any] (multitasking: Bool) is
   par Read, Grant_Read, Reject_Read, Write, Grant_Write, Reject_Write,
       Protection, Grant_Protection, Reject_Protection
   in
      SOURCE [...] (ip1, secure, privileged, data1, multitasking)
   || SOURCE [...] (ip2, secure, privileged, data2, multitasking)
   || SOURCE [...] (ip3, secure, non_privileged, data1, multitasking)
   || SOURCE [...] (ip4, secure, non_privileged, data2, multitasking)
   || SOURCE [...] (ip5, non_secure, privileged, data1, multitasking)
   || SOURCE [...] (ip6, non_secure, privileged, data2, multitasking)
   || SOURCE [...] (ip7, non_secure, non_privileged, data1, multitasking)
   || SOURCE [...] (ip8, non_secure, non_privileged, data2, multitasking)
   || TARGET [...] (ip0)
   end par
end process

------------------------------------------------------------------------------

process SOC_2 [Read, Grant_Read, Reject_Read, Write, Grant_Write, Reject_Write,
               Protection, Grant_Protection, Reject_Protection: Bus,
               Change_Source_Config: any] is
   par Read, Grant_Read, Reject_Read, Write, Grant_Write, Reject_Write,
       Protection, Grant_Protection, Reject_Protection
   in
      SOURCE [...] (ip1, secure, privileged, data1, true)
   || TARGET [...] (ip0)
   end par
end process

------------------------------------------------------------------------------

process MAIN [Read, Grant_Read, Reject_Read, Write, Grant_Write, Reject_Write,
              Protection, Grant_Protection, Reject_Protection: Bus,
              Change_Source_Config: any] is
   SOC [...] (false)
end process

------------------------------------------------------------------------------

process PURPOSE_1 [Reject_Read, Reject_Write, Reject_Protection,
                   TESTOR_ACCEPT: none] is
   -- any reject
   select
      Reject_Read
   [] Reject_Write
   [] Reject_Protection
   end select;
   loop TESTOR_ACCEPT end loop
end process

------------------------------------------------------------------------------

process PURPOSE_2 [Grant_Read, Grant_Write, Grant_Protection, Reject_Read,
                   Reject_Write, Reject_Protection, TESTOR_ACCEPT: none] is
   -- any transaction (all possible outcomes) in any order
   par
      Grant_Read
   || Grant_Write
   || Grant_Protection
   || Reject_Read
   || Reject_Write
   || Reject_Protection
   end par;
   loop TESTOR_ACCEPT end loop
end process

------------------------------------------------------------------------------

process PURPOSE_3 [Grant_Read, Grant_Write, Grant_Protection, Reject_Read,
                   Reject_Write, Reject_Protection, TESTOR_ACCEPT: none] is
   -- any transaction (all possible outcomes) in a seqential order
   Grant_Read;
   Grant_Write;
   Grant_Protection;
   Reject_Read;
   Reject_Write;
   Reject_Protection;
   loop TESTOR_ACCEPT end loop
end process

------------------------------------------------------------------------------

process PURPOSE_4 [Read, Grant_Read, Write, Grant_Protection: Bus,
                   TESTOR_ACCEPT, TESTOR_REFUSE: none] is
   -- granted read with different security/privilege than the preceding write
   var s,t: security, p,q: privilege, d: data in
      Grant_Protection (?any ip, ip0, ?s, ?p);
      Write (?any ip, ip0, s, p, ?d);
      -- forbid any change of the security/privilege of the target
      select
         null
      [] Grant_Protection (?any ip, ip0, ?s, ?p);
         loop TESTOR_REFUSE end loop
      end select;
      Read (?any ip, ip0, ?t, ?q) where (s != t) or (p != q);
      Grant_Read (?any ip, ip0, d);
      loop TESTOR_ACCEPT end loop
   end var
end process

------------------------------------------------------------------------------

end module
\end{lstlisting}

\section{PSS Model}
\label{app:pss}

\begin{lstlisting}[language=PSS,lastline=542]
component pss_top {

  // -------------------------------------------------------------------------
  // Types
  // -------------------------------------------------------------------------
  
  enum data_e {
    data1, data2
  }
  
  enum security_e {
    secure, non_secure
  }

  enum privilege_e {
    privileged, non_privileged
  }

  // -------------------------------------------------------------------------
  // Stream Flow Objects for Communication
  // (three streams per operation, for request, grant, and reject)
  // -------------------------------------------------------------------------

  // streams for read

  stream request_read_stream {
    rand security_e  sec;  // security of the source requesting to read
    rand privilege_e priv; // privilege of the source requesting to read
  }
  pool request_read_stream request_read_stream_pool;
  bind request_read_stream_pool *;

  stream grant_read_stream {
    rand data_e data; // read data
  }
  pool grant_read_stream grant_read_stream_pool;
  bind grant_read_stream_pool *;
  
  stream reject_read_stream {}
  pool reject_read_stream reject_read_stream_pool;
  bind reject_read_stream_pool *;
  
  // streams for write

  stream request_write_stream {
    rand security_e  sec;  // security of the source requesting to write
    rand privilege_e priv; // privilege of the source requesting to write
    rand data_e      data; // data to be written
  }
  pool request_write_stream request_write_stream_pool;
  bind request_write_stream_pool *;
  
  stream grant_write_stream {}
  pool grant_write_stream grant_write_stream_pool;
  bind grant_write_stream_pool *;

  stream reject_write_stream {}
  pool reject_write_stream reject_write_stream_pool;
  bind reject_write_stream_pool *;
  
  // streams for setting the protection

  stream request_protection_stream {
    rand security_e  sec;       // security of the requesting source
    rand privilege_e priv;      // privilege of the requesting source
    rand security_e  next_sec;  // new security
    rand privilege_e next_priv; // new privilege
  }
  pool request_protection_stream request_protection_stream_pool;
  bind request_protection_stream_pool *;
  
  stream grant_protection_stream {}
  pool grant_protection_stream grant_protection_stream_pool;
  bind grant_protection_stream_pool *;

  stream reject_protection_stream {}
  pool reject_protection_stream reject_protection_stream_pool;
  bind reject_protection_stream_pool *;
  
  // -------------------------------------------------------------------------
  // Finite State Machine for the Source
  // -------------------------------------------------------------------------

  enum source_state_e {
    idle, read, write, change
  }
  
  state source_state {
    rand source_state_e sstate;
    rand data_e      data;
    rand security_e  sec;
    rand privilege_e priv;
  }
  pool source_state source_state_pool;
  bind source_state_pool *;

  // initialize source
  action init_source {
    input  source_state in_state;
    output source_state out_state;

    constraint in_state.initial == true;
    // fix initial values (to cut nondeterminism)
    constraint out_state.sstate == idle;
    constraint out_state.sec    == secure;
    constraint out_state.priv   == privileged;
    constraint out_state.data   == data1;
  }

  // source read request
  action s_request_read {
    input  source_state        in_state;
    output source_state        out_state;
    output request_read_stream out_stream;

    constraint in_state.initial == false;
    // idle -> read
    constraint in_state.sstate  == idle;
    constraint out_state.sstate == read;
    // maintain fields
    constraint out_state.sec   == in_state.sec;
    constraint out_state.priv  == in_state.priv;
    constraint out_state.data  == in_state.data;
    // write to stream
    constraint out_stream.sec  == in_state.sec;
    constraint out_stream.priv == in_state.priv;
  }

  action s_grant_read {
    input  source_state      in_state;
    input  grant_read_stream in_stream;
    output source_state      out_state;

    constraint in_state.initial == false;
    // read -> idle
    constraint in_state.sstate  == read;
    constraint out_state.sstate == idle;
    // maintain fields
    constraint out_state.sec   == in_state.sec;
    constraint out_state.priv   == in_state.priv;
    constraint out_state.data   == in_state.data;
    // no constraint on the (thus, random) data read from the input stream
  }

  action s_reject_read {
    input  source_state       in_state;
    input  reject_read_stream in_stream;
    output source_state       out_state;

    constraint in_state.initial == false;
    // read -> idle
    constraint in_state.sstate  == read;
    constraint out_state.sstate == idle;
    // maintain fields
    constraint out_state.sec    == in_state.sec;
    constraint out_state.priv   == in_state.priv;
    constraint out_state.data   == in_state.data;
  }

  // source write request
  action s_request_write {
    input  source_state         in_state;
    output source_state         out_state;
    output request_write_stream out_stream;

    constraint in_state.initial == false;
    // idle -> write
    constraint in_state.sstate  == idle;
    constraint out_state.sstate == write;
    // maintain fields
    constraint out_state.sec    == in_state.sec;
    constraint out_state.priv   == in_state.priv;
    constraint out_state.data   == in_state.data;
    // write to stream
    constraint out_stream.data == in_state.data;
    constraint out_stream.sec  == in_state.sec;
    constraint out_stream.priv == in_state.priv;
  }

  action s_grant_write {
    input  source_state in_state;
    input  grant_write_stream in_stream;
    output source_state out_state;

    constraint in_state.initial == false;
    // write -> idle
    constraint in_state.sstate  == write;
    constraint out_state.sstate == idle;
    // maintain fields
    constraint out_state.sec    == in_state.sec;
    constraint out_state.priv   == in_state.priv;
    constraint out_state.data   == in_state.data;
  }

  action s_reject_write {
    input  source_state        in_state;
    input  reject_write_stream in_stream;
    output source_state        out_state;

    constraint in_state.initial == false;
    // write -> idle
    constraint in_state.sstate  == write;
    constraint out_state.sstate == idle;
    // maintain fields
    constraint out_state.sec    == in_state.sec;
    constraint out_state.priv   == in_state.priv;
    constraint out_state.data   == in_state.data;
  }

  // source protection change request
  action s_request_protection {
    input  source_state              in_state;
    output source_state              out_state;
    output request_protection_stream out_stream;

    constraint in_state.initial == false;
    // idle -> change
    constraint in_state.sstate  == idle;
    constraint out_state.sstate == change;
    // maintain fields
    constraint out_state.sec    == in_state.sec;
    constraint out_state.priv   == in_state.priv;
    constraint out_state.data   == in_state.data;
    // write to stream
    // no constraint on the new security and new privilege of the target
    // but source still states its security and privilege
    constraint out_stream.sec  == in_state.sec;
    constraint out_stream.priv == in_state.priv;
  }

  action s_grant_protection {
    input  source_state            in_state;
    input  grant_protection_stream in_stream;
    output source_state            out_state;

    constraint in_state.initial == false;
    // change -> idle
    constraint in_state.sstate  == change;
    constraint out_state.sstate == idle;
    // maintain fields
    constraint out_state.sec    == in_state.sec;
    constraint out_state.priv   == in_state.priv;
    constraint out_state.data   == in_state.data;
  }

  action s_reject_protection {
    input  source_state             in_state;
    input  reject_protection_stream in_stream;
    output source_state             out_state;

    constraint in_state.initial == false;
    // change -> idle
    constraint in_state.sstate  == change;
    constraint out_state.sstate == idle;
    // maintain fields
    constraint out_state.sec    == in_state.sec;
    constraint out_state.priv   == in_state.priv;
    constraint out_state.data   == in_state.data;
  }

  // change application running on the source: modify security, privilege, and data
  action change_source_config {
    input  source_state in_state;
    output source_state out_state;

    constraint in_state.initial == false;
    // stay in idle
    constraint in_state.sstate  == idle;
    constraint out_state.sstate == idle;
    // no constraint: randomly change source security, privilege, and data
    // (change application running on the source)
  }

  // -------------------------------------------------------------------------
  // Finite State Machine for the Target
  // -------------------------------------------------------------------------

  enum target_state_e {
    idle, read, write, change
  }
  
  state target_state { // Target FSM
    rand target_state_e sstate; // FSM STATE
    // Target internal data
    rand data_e data;           // current data
    rand security_e sec;        // current sec  protection
    rand privilege_e priv;      // current priv protection
    // Remember last transaction
    rand security_e tx_sec;     // transaction sec
    rand privilege_e tx_priv;   // transaction priv
    rand data_e tx_data;        // transaction data (write request)
    rand security_e next_sec;   // transaction change sec  request
    rand privilege_e next_priv; // transaction change priv request
  }
  pool target_state target_state_pool;
  bind target_state_pool *;

  action init_target {
    input  target_state in_state;
    output target_state out_state;

    constraint in_state.initial == true;
    // Cut nondeterminism by assigning inital values
    constraint out_state.sstate    == idle;
    constraint out_state.data      == data1;
    constraint out_state.sec       == non_secure;
    constraint out_state.priv      == non_privileged;
    constraint out_state.tx_sec    == non_secure;
    constraint out_state.tx_priv   == non_privileged;
    constraint out_state.tx_data   == data1;
    constraint out_state.next_sec  == non_secure;
    constraint out_state.next_priv == non_privileged;
  }

  // target READ request
  action t_request_read {
    input  target_state in_state;
    input  request_read_stream in_stream;
    output target_state out_state;

    constraint in_state.initial == false;
    // Idle -> Read
    constraint in_state.sstate  == idle;
    constraint out_state.sstate == read;
    // save stream data
    constraint out_state.tx_sec       == in_stream.sec;
    constraint out_state.tx_priv      == in_stream.priv;
    // Maintain fields
    constraint out_state.data      == in_state.data;
    constraint out_state.sec       == in_state.sec;
    constraint out_state.priv      == in_state.priv;
    constraint out_state.tx_data   == in_state.tx_data;
    constraint out_state.next_sec  == in_state.next_sec;
    constraint out_state.next_priv == in_state.next_priv;
    
  }

  action t_grant_read {
    input  target_state in_state;
    output target_state out_state;
    output grant_read_stream out_stream;

    constraint in_state.initial == false;
    // Read -> Idle
    constraint in_state.sstate  == read;
    constraint out_state.sstate == idle;
    // Maintain fields
    constraint out_state.data      == in_state.data;
    constraint out_state.sec       == in_state.sec;
    constraint out_state.priv      == in_state.priv;
    constraint out_state.tx_sec    == in_state.tx_sec;
    constraint out_state.tx_priv   == in_state.tx_priv;
    constraint out_state.tx_data   == in_state.tx_data;
    constraint out_state.next_sec  == in_state.next_sec;
    constraint out_state.next_priv == in_state.next_priv;
    // Check protection
    constraint (in_state.tx_sec == secure) ||
               (in_state.sec == non_secure);
    constraint (in_state.tx_priv == privileged) ||
               (in_state.priv == non_privileged);
    // Write on stream (give the data)
    constraint out_stream.data == in_state.data;
  }

  action t_reject_read {
    input  target_state in_state;
    output target_state out_state;
    output reject_read_stream out_stream;

    constraint in_state.initial == false;
    // Read -> Idle
    constraint in_state.sstate  == read;
    constraint out_state.sstate == idle;
    // Maintain fields
    constraint out_state.data      == in_state.data;
    constraint out_state.sec       == in_state.sec;
    constraint out_state.priv      == in_state.priv;
    constraint out_state.tx_sec    == in_state.tx_sec;
    constraint out_state.tx_priv   == in_state.tx_priv;
    constraint out_state.tx_data   == in_state.tx_data;
    constraint out_state.next_sec  == in_state.next_sec;
    constraint out_state.next_priv == in_state.next_priv;
    // Check protection
    constraint (in_state.sec == secure) || (in_state.priv == privileged);
    constraint (
                // sec check
                ((in_state.tx_sec  == non_secure) &&
                 (in_state.sec == secure))
                ||
                // priv check
                ((in_state.tx_priv == non_privileged) &&
                 (in_state.priv == privileged))
                );
    // Write on stream (fail verdict)
  }

  // target WRITE request
  action t_request_write {
    input  target_state in_state;
    input  request_write_stream in_stream;
    output target_state out_state;

    constraint in_state.initial == false;
    // Idle -> Write
    constraint in_state.sstate  == idle;
    constraint out_state.sstate == write;
    // save stream data
    constraint out_state.tx_sec       == in_stream.sec;
    constraint out_state.tx_priv      == in_stream.priv;
    constraint out_state.tx_data      == in_stream.data;
    // Maintain fields
    constraint out_state.data      == in_state.data;
    constraint out_state.sec       == in_state.sec;
    constraint out_state.priv      == in_state.priv;
    constraint out_state.next_sec  == in_state.next_sec;
    constraint out_state.next_priv == in_state.next_priv;
  }

  action t_grant_write {
    input  target_state in_state;
    output target_state out_state;
    output grant_write_stream out_stream;

    constraint in_state.initial == false;
    // write -> idle
    constraint in_state.sstate  == write;
    constraint out_state.sstate == idle;
    // Check protection
    constraint (in_state.tx_sec == secure) ||
               (in_state.sec == non_secure);
    constraint (in_state.tx_priv == privileged) ||
               (in_state.priv == non_privileged);
    // update data
    constraint out_state.data == in_state.tx_data;
    // maintain fields
    constraint out_state.sec       == in_state.sec;
    constraint out_state.priv      == in_state.priv;
    constraint out_state.tx_sec    == in_state.tx_sec;
    constraint out_state.tx_priv   == in_state.tx_priv;
    constraint out_state.tx_data   == in_state.tx_data;
    constraint out_state.next_sec  == in_state.next_sec;
    constraint out_state.next_priv == in_state.next_priv;
  }

  action t_reject_write {
    input  target_state in_state;
    output target_state out_state;
    output reject_write_stream out_stream;

    constraint in_state.initial == false;
    // Write -> Idle
    constraint in_state.sstate  == write;
    constraint out_state.sstate == idle;
    // Maintain fields
    constraint out_state.data      == in_state.data;
    constraint out_state.sec       == in_state.sec;
    constraint out_state.priv      == in_state.priv;
    constraint out_state.tx_sec    == in_state.tx_sec;
    constraint out_state.tx_priv   == in_state.tx_priv;
    constraint out_state.tx_data   == in_state.tx_data;
    constraint out_state.next_sec  == in_state.next_sec;
    constraint out_state.next_priv == in_state.next_priv;
    // Check protection
    constraint (in_state.sec == secure) || (in_state.priv == privileged);
    constraint (
                // sec check
                ((in_state.tx_sec  == non_secure) &&
                 (in_state.sec == secure))
                ||
                // priv check
                ((in_state.tx_priv == non_privileged) &&
                 (in_state.priv == privileged))
                );
    // Write on stream (fail verdict) 
  }

  // target Protection change request
  action t_request_protection {
    input  target_state in_state;
    input  request_protection_stream in_stream;
    output target_state out_state;
    
    constraint in_state.initial == false;
    // idle -> change
    constraint in_state.sstate  == idle;
    constraint out_state.sstate == change;
    // save stream data
    constraint out_state.tx_sec    == in_stream.sec;
    constraint out_state.tx_priv   == in_stream.priv;
    constraint out_state.next_sec  == in_stream.next_sec;
    constraint out_state.next_priv == in_stream.next_priv;
    // maintain fields
    constraint out_state.data      == in_state.data;
    constraint out_state.sec       == in_state.sec;
    constraint out_state.priv      == in_state.priv;
    constraint out_state.tx_data   == in_state.tx_data;
  }

  action t_grant_protection {
    input  target_state in_state;
    output target_state out_state;
    output grant_protection_stream out_stream;

    constraint in_state.initial == false;
    // Change protection -> Idle
    constraint in_state.sstate  == change;
    constraint out_state.sstate == idle;
    // Update protection
    constraint out_state.sec  == in_state.tx_sec;
    constraint out_state.priv == in_state.tx_priv;
    // Maintain fields
    constraint out_state.data      == in_state.data;
    constraint out_state.tx_sec    == in_state.tx_sec;
    constraint out_state.tx_priv   == in_state.tx_priv;
    constraint out_state.tx_data   == in_state.tx_data;
    constraint out_state.next_sec  == in_state.next_sec;
    constraint out_state.next_priv == in_state.next_priv;
    // Check protection
    constraint (in_state.tx_sec == secure);
    constraint (in_state.tx_priv == privileged);
  }

  action t_reject_protection {
    input  target_state in_state;
    output target_state out_state;
    output reject_protection_stream out_stream;

    constraint in_state.initial == false;
    // Change protection -> Idle
    constraint in_state.sstate  == change;
    constraint out_state.sstate == idle;
    // Maintain fields
    constraint out_state.data      == in_state.data;
    constraint out_state.sec       == in_state.sec;
    constraint out_state.priv      == in_state.priv;
    constraint out_state.tx_sec    == in_state.tx_sec;
    constraint out_state.tx_priv   == in_state.tx_priv;
    constraint out_state.tx_data   == in_state.tx_data;
    constraint out_state.next_sec  == in_state.next_sec;
    constraint out_state.next_priv == in_state.next_priv;
    // Check protection
    constraint (
\end{lstlisting}
\lstinline+}+\\

\section{Monolithic PSS Model}
\label{app:mono}

This PSS model also includes the four verification intents mentioned in Section~\ref{sec:test}.

\lstinputlisting[language=PSS]{PSS/RI_monolithic.pss}

\section{SVL Script for all Verification Steps}
\label{app:svl}

The following SVL script\footnote{SVL (Script Verification Language) is the language for describing verification scenarios for the CADP toolbox.} generates and compares the LTSs mentioned in Sections~\ref{sec:modeling:lnt} and \ref{sec:modeling:pss}.
It requires the translation to LNT of the monolithic PSS model (available in the MARS model repository).

\begin{lstlisting}[language=SVL,breaklines=true,lastline=77]
-- generation of the LTS for a SoC with 8 source IPs

"model_8_1.bcg" =
   reduction of "model_8_1.lnt";

-- generation of the LTS for a SoC with a single source IP

"model_2_1.bcg" =
   reduction of
   -- remove all actions from absent sources (only IP1 is present)
   total cut all but "[^!]* !IP1 .*" in
   "model_8_1.lnt":"SOC_2";

-- generation of the LTS for the PSS model

"RI_monolithic.bcg" =
   strong reduction of
   weak trace reduction of
   branching reduction of
   -- 4. suppression of supperfluous offers (to be completed)
   total rename
      "\(CHANGE_SOURCE_CONFIG\) .* \(![^!]* ![^!]* ![^!]*\) ![^!]* ![^!]* ![^!]* ![^!]* ![^!]*" -> "\1 \2",
      "\(GRANT_READ\) .* \(![^!]*\) ![^!]* ![^!]*" -> "\1 \2",
      "\(GRANT_WRITE\).*" -> "\1",
      "\(GRANT_PROTECTION\) .* \(![^!]* ![^!]*\) ![^!]* ![^!]* ![^!]*" -> "\1 \2",
      "\(REJECT_[A-Z]*\).*" -> "\1",
      "REQUEST_\(READ\) ![^!]* \(![^!]* ![^!]*\) .*" -> "\1 \2",
      "REQUEST_\(WRITE\) ![^!]* \(![^!]* ![^!]* ![^!]*\) .*" -> "\1 \2",
      "REQUEST_\(PROTECTION\) ![^!]* \(![^!]* ![^!]*\) .* \(![^!]* ![^!]*\)" -> "\1 \2 \3"
   in
   -- 3. suppression of the prefix SOURCE/TARGET
   rename
      "SOURCE_\([A-Z_]*\)" -> "\1",
      "TARGET_\([A-Z_]*\)" -> "\1"
   in
   -- 2. removal of the first offer (indicating the action)
   total rename
      "\([^!]*\)!PSS_TOP_X_[^!]*\(.*\)" -> "\1\2"
   in
   -- 1. removal of the gate prefix "PSS\_TOP\_X\_"
   rename
      "PSS_TOP_X_\(.*\)" -> "\1"
   in
   -- hiding initialisation and interaction with the state flow object
   divbranching reduction of
   hide
      ".*OUTPUT",
      ".*INPUT",
      ".*INIT_SYSTEM"
   in
      "../PSS/RI_monolithic.lnt"
   end hide;

-- comparion of the three LTSs

property MODEL_EQUIVALENCE
   "after hiding ip identities, alls models are equivalent"
is
   branching comparison
      hide CHANGE_SOURCE_CONFIG in
      total rename "\([^ ]*\) ![^!]* ![^!]*\(.*\)" -> "\1 \2" in
      "model_8_1.bcg"
      ==
      hide CHANGE_SOURCE_CONFIG in
      total rename "\([^ ]*\) ![^!]* ![^!]*\(.*\)" -> "\1 \2" in
      "model_2_1.bcg";
   expected TRUE;

   branching comparison
      hide CHANGE_SOURCE_CONFIG in
      total rename "\([^ ]*\) ![^!]* ![^!]*\(.*\)" -> "\1 \2" in
      "model_8_1.bcg"
      ==
      hide CHANGE_SOURCE_CONFIG in
      "RI_monolithic.bcg";
   expected TRUE;
end property
\end{lstlisting}

\end{document}